\pdfoutput=1

\documentclass[onecolumn]{emulateapj}

\usepackage{amsmath}
\usepackage{graphicx}
\usepackage{color}
\usepackage[colorlinks]{hyperref}
\hypersetup{
    colorlinks,	
    citecolor=blue,
}

\newcommand{\be}{\begin{eqnarray}}
\newcommand{\ee}{\end{eqnarray}}

\def\jcap{JCAP}

\shorttitle{Testing the Kerr Black Hole Hypothesis with GX~339--4}
\shortauthors{Tripathi et al.}

\begin{document}

\title{Testing the Kerr Black Hole Hypothesis with GX~339--4\\by a combined analysis of its thermal spectrum and reflection features}

\author{Ashutosh~Tripathi\altaffilmark{1}, Askar~B.~Abdikamalov\altaffilmark{1,2}, Dimitry~Ayzenberg\altaffilmark{3}, Cosimo~Bambi\altaffilmark{1,\dag}, Victoria~Grinberg\altaffilmark{4}, and Menglei~Zhou\altaffilmark{4}}

\altaffiltext{1}{Center for Field Theory and Particle Physics and Department of Physics, Fudan University, 200438 Shanghai, China. \email[\dag E-mail: ]{bambi@fudan.edu.cn}}
\altaffiltext{2}{Ulugh Beg Astronomical Institute, Tashkent 100052, Uzbekistan}
\altaffiltext{3}{Theoretical Astrophysics, Eberhard-Karls Universit\"at T\"ubingen, D-72076 T\"ubingen, Germany}
\altaffiltext{4}{Institut f\"ur Astronomie und Astrophysik, Eberhard-Karls Universit\"at T\"ubingen, D-72076 T\"ubingen, Germany}

\begin{abstract}
We analyze simultaneous observations with \textsl{NuSTAR} and \textsl{Swift} of the black hole binary GX~339--4 in which we clearly detect both a strong thermal component and prominent relativistic reflection features. We employ {\tt nkbb} and {\tt relxill\_nk}, which are the state of the art of, respectively, the thermal and reflection models to test the Kerr black hole hypothesis. We obtain unprecedented precise measurements of the Johannsen deformation parameter $\alpha_{13}$: $\alpha_{13} = -0.012_{-0.039}^{+0.011}$ (lamppost coronal geometry) and $\alpha_{13} = -0.010_{-0.018}^{+0.024}$ (broken power-law emissivity profile) at a 90\% confidence level, where the Kerr metric corresponds to $\alpha_{13} = 0$. We investigate the systematic uncertainties by fitting the data with different models. 
\end{abstract}


%%%%%%%%%%%%%%%%%%%%%%%%%%%%%%%

\section{Introduction}

In 4-dimensional General Relativity, uncharged black holes are described by the Kerr solution, which is completely specified by two parameters associated, respectively, to the mass $M$ and the spin angular momentum $J$ of the compact object~\citep{1971PhRvL..26..331C,1975PhRvL..34..905C}. The spacetime metric around astrophysical black holes is thought to be described well by the Kerr solution, as the presence of an accretion disk, nearby stars, or a possible non-vanishing electric charge of the black hole are normally negligible for the strong gravitational field near the event horizon~\citep[see, e.g.,][and references therein]{2009JCAP...09..013B,2018AnP...53000430B}. In such a framework, the continuum-fitting method~\citep{1997ApJ...482L..155C,2014SSRv..183..295M} and X-ray reflection spectroscopy~\citep{1989MNRAS.238..729F,2006ApJ...652.1028B,2019NatAs...3...41R} are the two leading techniques for measuring the spins of accreting black holes.

The continuum-fitting method refers to the analysis of the thermal spectrum of geometrically thin and optically thick accretion disks~\citep{1997ApJ...482L..155C,2014SSRv..183..295M}. The model normally assumes a Novikov-Thorne disk with inner edge at the innermost stable circular orbit (ISCO)~\citep{1973blho.conf..343N,1974ApJ...191..499P}, which requires to select sources in the soft state and with an Eddington-scaled disk luminosity between $\sim$5\% to $\sim$30\%~\citep{2010ApJ...718L.117S,2011MNRAS.414.1183K,2014SSRv..183..295M}. The spectrum depends on five parameters: the black hole mass $M$, the distance $D$, the inclination angle of the disk with respect to the line of sight of the distant observer $i$, the black hole spin parameter $a_* = J/M^2$ (here and in the rest of the paper, we employ natural units in which $G_{\rm N} = c = 1$), and the mass accretion rate $\dot{M}$. In general, the spectrum is degenerate with respect to these five parameters and it is not possible to infer all their values by fitting the data. However, if we can get independent estimates of $M$, $D$, and $i$, for example from optical observations, then we can fit the thermal component of the disk to measure $a_*$ and $\dot{M}$.

X-ray reflection spectroscopy is the analysis of relativistic reflection features generated by illumination of a cold accretion disk by a hot corona~\citep{1989MNRAS.238..729F,2006ApJ...652.1028B,2019NatAs...3...41R}. In the rest-frame of the gas, the reflection spectrum is characterized by some narrow fluorescent emission lines in the soft X-ray band, in particular the iron K$\alpha$ complex at 6.4-6.97~keV, depending on the ionization of the iron ions, and the Compton hump peaked at 20-30~keV~\citep{1995MNRAS.273..837M,2005MNRAS.358..211R,2010ApJ...718..695G}. The reflection spectrum of the whole disk as seen by a distant observer is the result of the combination of radiation emitted from different parts of the accretion disk and altered by relativistic effects (Doppler boosting, gravitational redshift, and light bending) of different intensity. In the presence of the correct astrophysical model and high quality data, the analysis of these relativistically distorted reflection features can tell us about the morphology of the accretion flow and the properties of the strong gravity region near the black hole event horizon.

Motivated by a number of scenarios predicting the possibility that the spacetime metric around astrophysical black holes presents macroscopic deviations from the Kerr metric~\citep[see, e.g.,][]{2013PhLB..719..419D,2014PhRvL.112v1101H,2014PhRvD..90l4033G,2017NatAs...1E..67G}, the continuum-fitting method and X-ray reflection spectroscopy can be readapted to test the Kerr black hole hypothesis~\citep[early work in this direction is, e.g., in][]{2003IJMPD..12...63L,2009GReGr..41.1795S,2009PhRvD..79f4001H,2009PhRvD..80d4021H,2011ApJ...731..121B,2012ApJ...761..174B,2013ApJ...773...57J,2013PhRvD..87b3007B}. For a review, see, e.g., \citet{2016CQGra..33f4001B,2016CQGra..33l4001J,2017RvMP...89b5001B,2018GReGr..50..100K}.

Some of us have recently developed two {\tt xspec} models~\citep{xspec} to test the Kerr black hole hypothesis and they represent the state of the art in the field. {\tt nkbb} is designed to calculate the thermal spectrum of geometrically thin and optically thick accretion disks in non-Kerr spacetimes~\citep{2019PhRvD..99j4031Z}. {\tt relxill\_nk}~\citep{2017ApJ...842...76B,2019ApJ...878...91A} is an extension of the {\tt relxill} package~\citep{2013MNRAS.430.1694D,2014ApJ...782...76G} to analyze reflection features of thin accretion disks in non-Kerr spacetimes. Most of our past studies have followed the so-called bottom-up approach: the models employ a parametric black hole spacetime in which deviations from the Kerr solution are parametrized by a number of {\it deformation parameters}. The Kerr solution is recovered when all deformation parameters vanish. From the comparison of theoretical predictions with observational data, we can estimate the values of these deformation parameters and check whether their measurement is consistent with zero, which is the value required to recover the Kerr metric and thus predicted by General Relativity~\citep[see, e.g.,][]{2018PhRvL.120e1101C,2019ApJ...884..147Z,2019ApJ...875...56T,2020ApJ...897...84T}.

In this paper, we present our analysis of two simultaneous observations of \textsl{NuSTAR} and \textsl{Swift} of the black hole binary GX~339--4. These observations were previously studied by~\citet{2016ApJ...821L...6P} in the framework of General Relativity. Since the spectrum clearly shows both a strong thermal component and prominent relativistic reflection features, we have the possibility of using {\tt nkbb} and {\tt relxill\_nk} at the same time. This is the first time that we can test the Kerr metric from the combined analysis of thermal and reflection spectra and, together with the high quality data of these \textsl{NuSTAR} and \textsl{Swift} observations of GX~339--4, we are able to get quite a precise measurement of the deformation parameter $\alpha_{13}$ of the Johannsen metric~\citep{2013PhRvD..88d4002J}.

The content of the paper is as follows. In Section~\ref{s-obs}, we describe the \textsl{NuSTAR} and \textsl{Swift} observations and our data reduction. In Section~\ref{s-ana}, we present our spectral analysis with {\tt nkbb} and {\tt relxill\_nk}. In Section~\ref{s-dis}, we discuss our results and we try to estimate some systematic uncertainties related to our theoretical model. Our conclusions are reported in Section~\ref{s-con}.

%%%%%%%%%%%%%%%%%%%%%%%%%%%%%%%

\begin{table*}
\centering
{\renewcommand{\arraystretch}{1.2}
\begin{tabular}{lcc}
\hline\hline
Mission &  Observation~ID & Exposure (ks) \\
\hline
\textsl{NuSTAR} & 80001015003 & 30 \\
\textsl{Swift} & 00081429002 & 2 \\
\hline\hline
\end{tabular}
}
%\vspace{0.3cm}
\caption{\rm Observations analyzed in the present work. \label{t-obs}}
\end{table*}

\section{Observation and data reduction}\label{s-obs}

GX~339--4 is a very active low-mass X-ray binary with an outburst every 1-3~years. It was discovered in 1973 by \textsl{OSO--7}~\citep{1973ApJ...184L..67M}. Since it often enters outbursts, there are many observations of this source and with different X-ray missions. It was simultaneously observed by \textsl{NuSTAR} and \textsl{Swift} on 2015 March 11, after \textsl{Swift} monitoring had detected strong thermal and power-law components in its spectrum on 2015 March 4. Observation IDs and exposure times of the two missions are reported in Tab.~\ref{t-obs}. The analysis of these data were presented in \citet{2016ApJ...821L...6P} (assuming that the spacetime metric around the black hole is described by the Kerr solution) to estimate the values of the black hole spin parameter $a_*$, black hole mass $M$, and distance of the source $D$. Our study follows the analysis presented in \citet{2016ApJ...821L...6P} but we employ the non-Kerr models {\tt nkbb} and {\tt relxill\_nk} in order to measure the deformation parameter $\alpha_{13}$ of the Johannsen metric~\citep{2013PhRvD..88d4002J}.
The line element of the Johannsen spacetime is reported in Appendix, where we also list the main properties of this black hole metric.

The XRT/\textsl{Swift} data are processed to the cleaned event files using the {\tt xrtpipeline} module of the Swift XRT data analysis software (SWXRTDAS) v3.5.0, distributed as part of the {\tt heasoft} spectral analysis package v6.28 and CALDB version 20200724. These cleaned events are further processed and filtered using {\tt xselect} v2.4g. The data are affected by pile-up. To avoid the pile-up region during the spectra extraction, an annulus with inner radius of 25~arcsec and outer radius of 45~arcsec is taken as source region around the center of the source. We use the spectral redistribution matrices files (RMF) as given in CALDB. Ancillary
files (ARF) are generated using the module {\tt xrtmkarf}. We bin the data to attain a signal-to-noise ratio
of 30 and we use the data in the range of 1-4~keV for further analysis.

Simultaneous to the \textsl{Swift} data, we have a 30~ks \textsl{NuSTAR} observation (Obs. ID 80001015003). The raw data is first converted into cleaned event files using the module {\tt nupipeline}, which is the part of \textsl{NuSTAR} data analysis software (NuSTARDAS) v2.0.0 and CALDB version 20200912. The source region is extracted as a circular region of 150~arcsec centered on the source. The background region of 100~arcsec is selected far from the source in order to minimize the contamination from source photons. The spectra and the RMF and ARF files are extracted using {\tt nuproducts}. We bin the FPMA and FPMB spectra in order to reach a signal-to-noise ratio of 50. We use the 3-60~keV band for further analysis.

For a quick comparison with the results in \citet{2016ApJ...821L...6P}, in the rest of the paper we report the 1-$\sigma$ confidence level uncertainties, unless stated otherwise. We adopt {\it wilms} abundances~\citep{2000ApJ...542..914W} and {\it vern} cross-sections~\citep{1996ApJ...465..487V}.

%%%%%%%%%%%%%%%%%%%%%%%%%%%%%%%

\begin{figure}[t]
\begin{center}
\includegraphics[width=8.5cm,trim={1.5cm 0.5cm 3.5cm 5.0cm},clip]{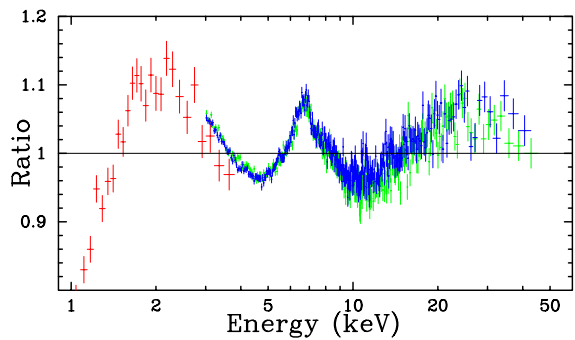}
\end{center}
\vspace{-0.5cm}
\caption{Data to best-fit model ratio for an absorbed power law + disk blackbody spectrum [in {\sc xspec} language, {\tt tbabs*(diskbb + powerlaw)}]. Red crosses are for XRT/\textsl{Swift} data, green crosses for FPMA/\textsl{NuSTAR} data, and blue crosses for FPMB/\textsl{NuSTAR} data. \label{f-ratio}}
\vspace{0.5cm}
\end{figure}

\section{Spectral analysis}\label{s-ana}

If we fit the \textsl{NuSTAR}+\textsl{Swift} data with an absorbed disk multi-temperature blackbody spectrum + power-law component (Model~0) -- {\tt tbabs*(diskbb + powerlaw)} in {\tt xspec} language\footnote{We employ standard notation, where {\tt *} indicates either multiplication or convolution, and {\tt +} indicates summation. {\tt tbabs} describes the Galactic absorption~\citep{2000ApJ...542..914W}. {\tt diskbb} describes the multi-temperature blackbody spectrum of the disk~\citep{1984PASJ...36..741M}. {\tt powerlaw} describes the power-law component from the corona.} -- we find the residuals shown in Fig.~\ref{f-ratio}. The \textsl{NuSTAR} data clearly present a broad iron line around 7~keV and a Compton hump with a peak at 20-30~keV. The \textsl{Swift} data instead present a bump around 2~keV. This agrees with Fig.~1 in \citet{2016ApJ...821L...6P}, where the 2~keV bump is interpreted as a deficiency of the simple phenomenological model {\tt tbabs*(diskbb + powerlaw)}.

We add a relativistic reflection component with {\tt relxill\_nk}~\citep{2017ApJ...842...76B,2019ApJ...878...91A} and we replace the Newtonian multi-temperature blackbody model {\tt diskbb} with {\tt nkbb}~\citep{2019PhRvD..99j4031Z}. Initially we consider two models:

\vspace{0.15cm}

Model~A: {\tt tbabs*simplcutx*(nkbb + relxill\_nk)}

\vspace{0.15cm}

Model~B: {\tt tbabs*(nkbb + comptt + relxill\_nk)}

\vspace{0.15cm}

In Model~A, {\tt simplcutx} describes the Comptonization of the corona~\citep{2017ApJ...836..119S}, so {\tt simplcutx*nkbb} gives as output both the thermal spectrum of the accretion disk and the Comptonized photons. In principle, even the reflection spectrum can be Comptonized too, so we have {\tt simplcutx*relxill\_nk}. In Model~B, we use {\tt comptt} to describes the Comptonized thermal photons from the disk~\citep{1994ApJ...434..570T} and we neglect the Comptonization of the reflection spectrum, which would correspond to the case of a very compact corona that does not cover well the inner part of the accretion disk.

We note that the values of the inclination angle of the disk $i$, of the black hole spin parameter $a_*$, and of the Johannsen deformation parameter $\alpha_{13}$ are tied between {\tt nkbb} and {\tt relxill\_nk}. The photon index $\Gamma$ in {\tt simplcutx} is tied to that in {\tt relxill\_nk} and the coronal temperature in {\tt simplcutx} and in {\tt comptt} is related to the high energy cut-off in {\tt relxill\_nk} by $2 k T_{\rm e} = E_{\rm cut}$. In {\tt relxill\_nk}, we consider two emissivity profiles: the lamppost coronal geometry ({\tt relxilllp\_nk} with one free parameter for the emissivity profile, the coronal height $h$) and the broken power-law profile (the actual {\tt relxill\_nk} model with three free parameters: inner emissivity index $q_{\rm in}$, outer emissivity index $q_{\rm out}$, and breaking radius $R_{\rm br}$). The black hole mass $M$ and the black hole distance $D$ in {\tt nkbb} are free in the fit: this is not the standard approach of the continuum-fitting method, where $M$, $D$, and $i$ are determined from other observations. However, here the idea is that $i$ and $a_*$ are mainly determined from the reflection spectrum, and the thermal component can thus constrain $M$ and $D$. We note that \citet{2016ApJ...821L...6P} can estimate $a_*$ and $i$ from the analysis of the \textsl{NuSTAR} data, and then, when they add the \textsl{Swift} data, they can get an estimate of $M$ and $D$. In our case, the fit is a bit more complicated. From the analysis of the \textsl{NuSTAR} data, if we do not freeze the column density $N_{\rm H}$ in {\tt tbabs}, we can only get weak constraints on $a_*$ and $\alpha_{13}$, because of the degeneracy between these two parameters. However, when we simultaneously analyze the \textsl{NuSTAR} and \textsl{Swift} data, we are able to constrain all parameters, both in Model~A and Model~B, and this proves the enhanced constraining power when we consistently fit two components that depend both on the same parameter(s) of the model.

Model~B provides a definitively better fit than Model~A, with a difference of $\chi^2$ around 250 for both the lamppost and the broken power-law models. \citet{2016ApJ...821L...6P} also try both models and arrive at the same conclusion, even if they do not mention the difference in $\chi^2$. In what follows, we will only consider Model~B.

\begin{figure*}[b]
\begin{center}
\includegraphics[width=17.0cm,trim={0cm 0cm 0cm 0cm},clip]{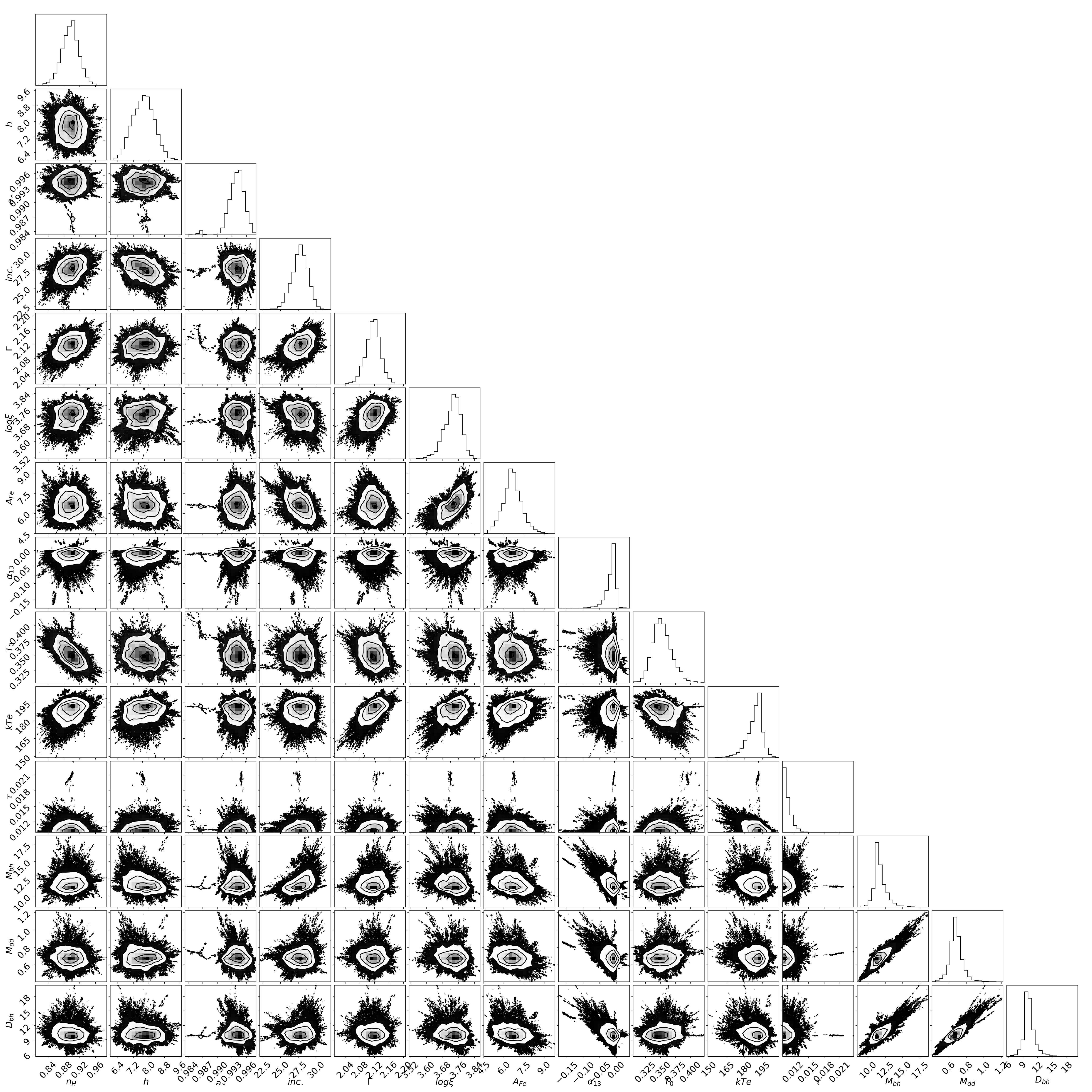}
\end{center}
\vspace{-0.2cm}
\caption{Corner plot for the free parameter-pairs (excluding calibration constants and component normalizations) in Model~B1 after the MCMC run. The 2D plots report the 1-, 2-, and 3-$\sigma$ confidence contours (corresponding, respectively, to $\Delta\chi^2 = 2.3$, 6.2, and 11.8). \label{f-mcmc1}}
\end{figure*}

\begin{figure*}[b]
\begin{center}
\includegraphics[width=17.0cm,trim={0cm 0cm 0cm 0cm},clip]{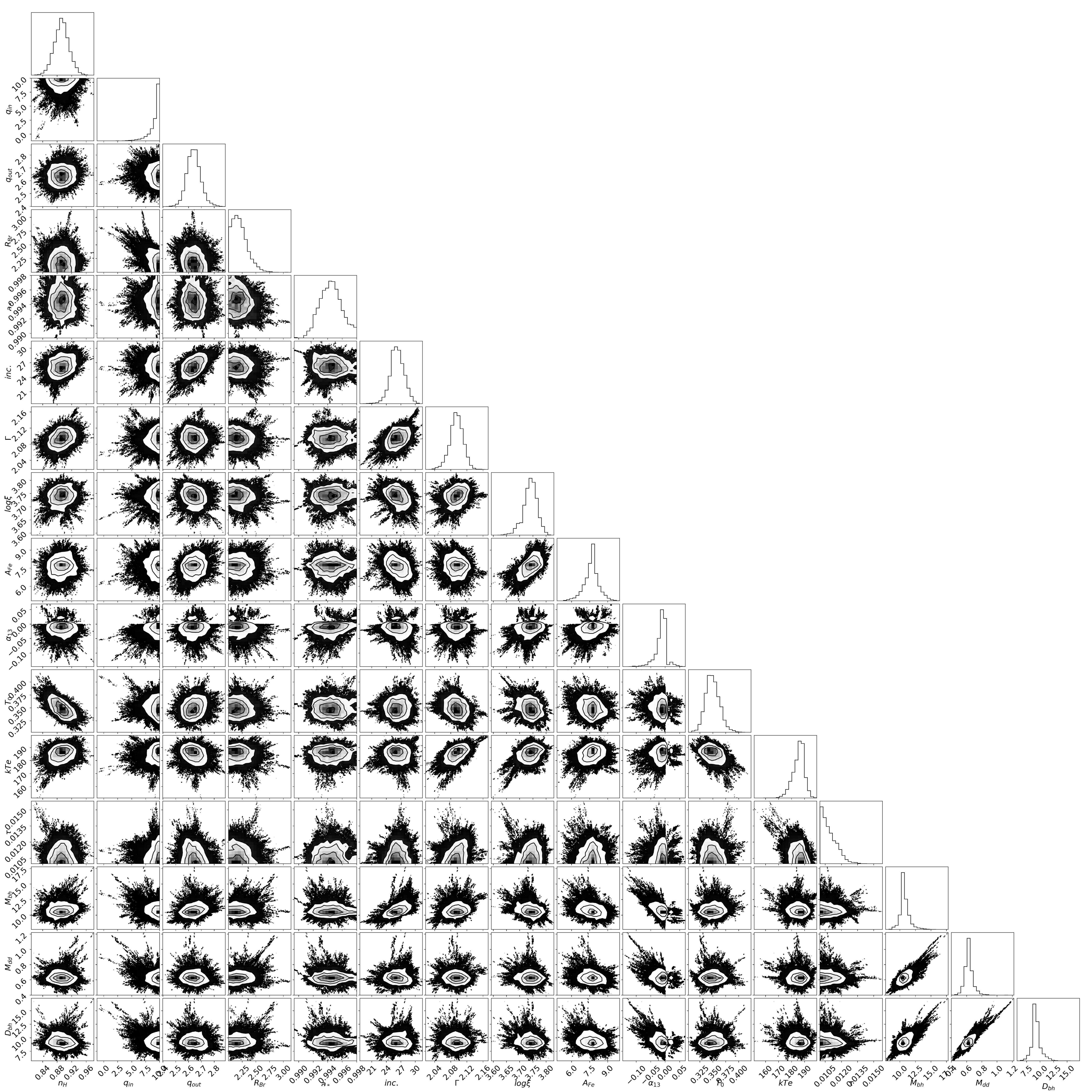}
\end{center}
\vspace{-0.2cm}
\caption{Corner plot for the free parameter-pairs (excluding calibration constants and component normalizations) in Model~B2 after the MCMC run. The 2D plots report the 1-, 2-, and 3-$\sigma$ confidence contours (corresponding, respectively, to $\Delta\chi^2 = 2.3$, 6.2, and 11.8). \label{f-mcmc2}}
\end{figure*}

\begin{figure*}[t]
\begin{center}
\includegraphics[width=8.5cm,trim={2.0cm 0.5cm 3.0cm 17.0cm},clip]{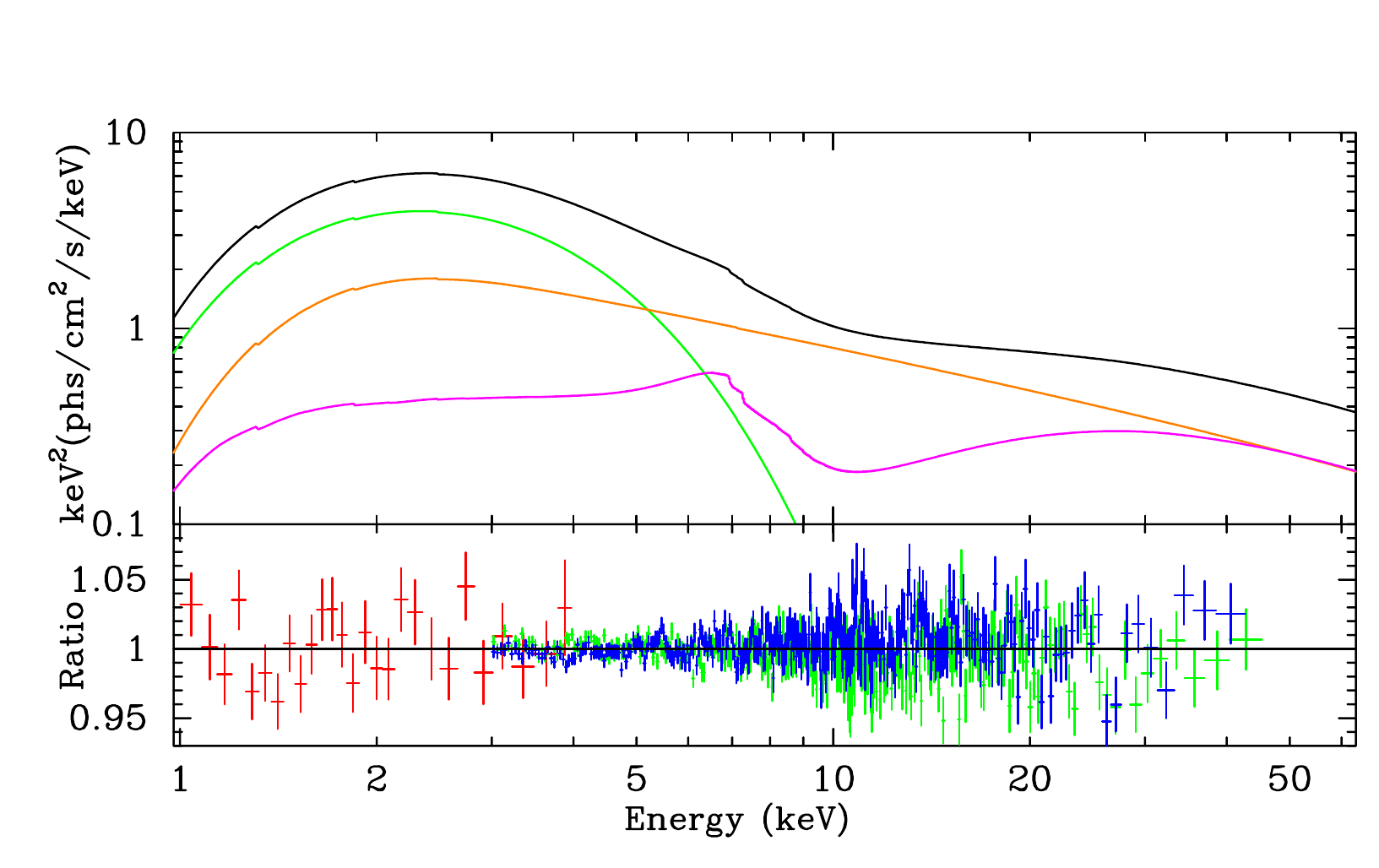}
\hspace{0.25cm}
\includegraphics[width=8.5cm,trim={2.0cm 0.5cm 3.0cm 17.0cm},clip]{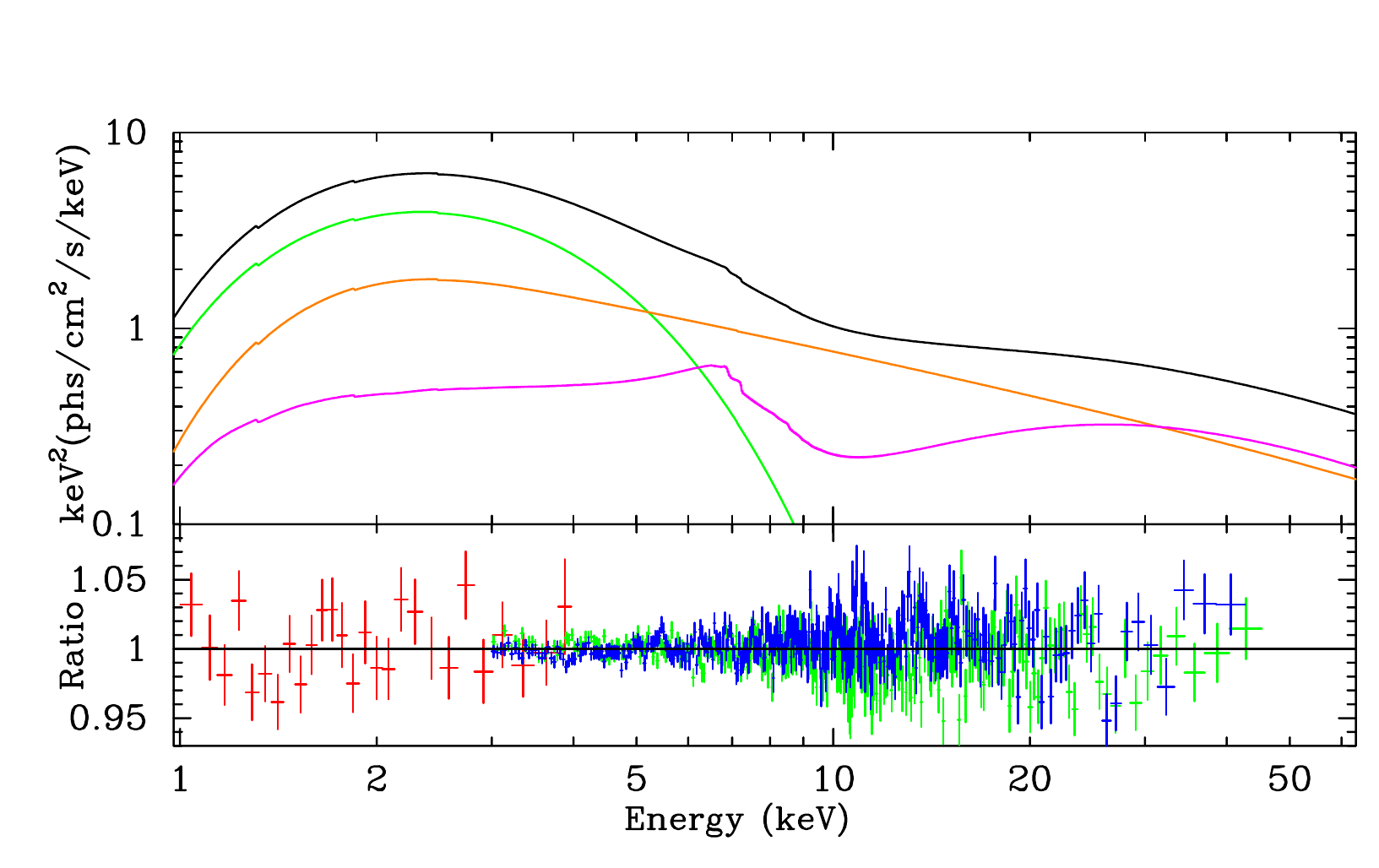}
\end{center}
\vspace{-0.5cm}
\caption{Best-fit model and data to best-fit model ratio for Model~B1 (left panel) and Model~B2 (right panel). The black curves are for the total model, the green curves for the thermal component ({\tt nkbb}), the orange curves for the continuum ({\tt comptt}), and the magenta curves for the relativistic reflection component ({\tt relxill\_nk}). In the ratio plots, red crosses are for XRT/\textsl{Swift} data, green crosses for FPMA/\textsl{NuSTAR} data, and blue crosses for FPMB/\textsl{NuSTAR} data. \label{f-mr}}
\vspace{0.5cm}
\end{figure*}

Since we have many free parameters in the fits, we perform a Markov Chain Monte-Carlo (MCMC) analysis of Model~B, for both the lamppost geometry (Model~B1) and the broken-power-law profile (Model~B2), using the python script by Jeremy Sanders which uses {\tt emcee} (MCMC Ensemble sampler implementing Goodman \& Weare algorithm)\footnote{Available on github at \href{https://github.com/jeremysanders/xspec_emcee}{https://github.com/jeremysanders/xspec\_emcee}.}. We use 500 walkers of 10,000~steps each, with an initial burn-in period of 1000~steps. This is 10~times the auto-correlation length, which is around 100. The best-fit values are shown in Tab.~\ref{t-fit} (second and third column). Fig.~\ref{f-mcmc1} and Fig.~\ref{f-mcmc2} show the corner plots with the 1- and 2-dimensional projections of the posterior probability distributions of the relevant free parameters for, respectively, Model~B1 and Model~B2. The best-fit models and the data to the best-fit model ratios are shown in Fig.~\ref{f-mr} (the difference between the two models can be hardly seen). The measurement of the Johannsen deformation parameter $\alpha_{13}$ is (90\% confidence level for one relevant parameter, $\Delta\chi^2 = 2.71$)
\be\label{eq-a13}
\alpha_{13} &=& -0.012_{-0.039}^{+0.011} \quad (\text{Model~B1}) \, , \nonumber\\
\alpha_{13} &=& -0.010_{-0.018}^{+0.024} \quad (\text{Model~B2}) \, .
\ee

\begin{table*}
\centering
{\renewcommand{\arraystretch}{1.2}
\begin{tabular}{lccccccc}
\hline\hline
Parameter &  Model~B1 & Model~B2 &  Model~B3 & Model~B4 &  Model~B5 & Model~B6 & Model~B7 \\
\hline
{\tt tbabs} && \\
$N_{\rm H}$ [$10^{22}$~cm$^{-2}$] & $0.90^{+0.02}_{-0.02}$& $0.89^{+0.02}_{-0.02}$ & $0.89^{+0.02}_{-0.02}$ & $0.87^{+0.02}_{-0.02}$ & $0.88^{+0.02}_{-0.02}$ & $0.93^{+0.02}_{-0.02}$ & $0.89^{+0.01}_{-0.01}$ \\
\hline
{\tt nkbb} && \\
$M$ [$M_\odot$] & $11.5^{+1.0}_{-0.5}$ & $10.6^{+0.7}_{-0.4}$ & $9.8^{+0.9}_{-0.6}$ & $12.7^{+1.4}_{-0.7}$ & $11.1^{+0.4}_{-0.5}$ & $11.9^{+0.7}_{-0.6}$ & $12.3^{+0.1}_{-0.1}$ \\
$\dot{M}$ [$10^{18}$~g/s] & $0.68^{+0.07}_{-0.06}$ & $0.63^{+0.05}_{-0.04}$ & $0.69^{+0.07}_{-0.06}$ & $0.70^{+0.09}_{-0.06}$ & $0.72^{+0.06}_{-0.05}$ & $0.65^{+0.06}_{-0.06}$ & $0.65^{+0.09}_{-0.06}$ \\
$D$ [kpc] & $10.1^{+1.2}_{-0.7}$ & $9.2^{+0.8}_{-0.5}$ & $10.4^{+1.3}_{-0.8}$ & $9.7^{+1.3}_{-0.7}$ & $9.8^{+0.5}_{-0.6}$ & $9.9^{+0.8}_{-0.7}$ & $9.7^{+0.2}_{-0.2}$ \\
$f_{\rm col}$ & 1.7$^\star$ & 1.7$^\star$ & 1.5$^\star$ & 1.9$^\star$ & 1.7$^\star$ & 1.7$^\star$ & 1.7$^\star$ \\ 
\hline
{\tt comptt} && \\
$T_0$ [keV] & $0.351^{+0.026}_{-0.019}$ & $0.349^{+0.019}_{-0.021}$ & $0.355^{+0.025}_{-0.014}$ & $0.36^{+0.03}_{-0.04}$ & $0.356^{+0.014}_{-0.015}$ & $0.351^{+0.015}_{-0.013}$ & $0.353^{+0.014}_{-0.013}$ \\
$T_{\rm plasma}$ [keV] & $190^{+4}_{-8}$& $186^{+3}_{-5}$ & $186^{+4}_{-6}$ & $189^{+5}_{-8}$ & $189^{+3}_{-4}$ & $214^{+4}_{-6}$ & $193^{+2}_{-2}$ \\
$\tau$ & $<0.01$& $<0.01$ & $<0.01$& $<0.01$ & $<0.01$& $<0.01$ & $<0.01$ \\
\hline
{\tt relxill\_nk} &&&&&&& \\
$R_{\rm in}$ [$r_{\rm g}$] & $-1^\star$ & $-1^\star$ & $-1^\star$ & $-1^\star$ & $1.9^{+0.6}_{-0.6}$ & $-1^\star$ & $-1^\star$ \\
$h$ [$r_{\rm g}$] & $7.7^{+0.5}_{-0.6}$ & -- & $7.6^{+0.6}_{-0.5}$ & $8.2^{+0.8}_{-0.7}$ & $7.6^{+0.7}_{-0.8}$& $8.0^{+0.5}_{-0.4}$ & $7.6^{+0.6}_{-0.5}$ \\
$q_{\rm in}$ & -- & $9.5^{+0.4}_{-1.3}$ & -- & -- & -- & -- & -- \\
$q_{\rm out}$ & -- & $2.64^{+0.05}_{-0.05}$ & -- & -- & -- & -- & -- \\
$R_{\rm br}$ [$r_{\rm g}$] & -- & $2.19^{+0.16}_{-0.12}$ & -- & -- & -- & -- & -- \\
$a_*$& $0.994^{+0.001}_{-0.002}$&$0.994^{+0.002}_{-0.001}$ & $0.994^{+0.002}_{-0.002}$ & $0.981^{+0.003}_{-0.003}$ & $0.981^{+0.004}_{-0.004}$ & $0.992^{+0.002}_{-0.002}$ & $0.989^{+0.006}_{-0.016}$\\
$i$ [deg] & $27.7^{+1.1}_{-1.2}$ & $26.2^{+1.5}_{-1.3}$ & $28.9^{+1.0}_{-1.2}$ & $27.6^{+1.2}_{-1.3}$ & $28.6^{+0.9}_{-1.0}$ & $26.3^{+1.0}_{-0.9}$ & $28.1^{+1.1}_{-1.0}$ \\
$\Gamma$ & $2.12^{+0.02}_{-0.02}$&$2.09^{+0.02}_{-0.02}$ & $2.11^{+0.02}_{-0.01}$ & $2.11^{+0.02}_{-0.02}$ & $2.11^{+0.02}_{-0.02}$ & $2.12^{+0.02}_{-0.02}$ & $2.13^{+0.02}_{-0.02}$ \\
$E_{\rm cut}$ [keV] & 2~$T_{\rm plasma}$ & 2~$T_{\rm plasma}$ & 2~$T_{\rm plasma}$ & 2~$T_{\rm plasma}$ & 2~$T_{\rm plasma}$ & $300^\star$ & 2~$T_{\rm plasma}$ \\ 
$\log\xi$ [erg~cm~s$^{-1}$] & $3.74^{+0.04}_{-0.05}$ & $3.74^{+0.02}_{-0.03}$ & $3.66^{+0.05}_{-0.05}$ & $3.71^{+0.04}_{-0.04}$ & $3.65^{+0.05}_{-0.04}$ & $3.49^{+0.03}_{-0.03}$ & $3.73^{+0.03}_{-0.03}$\\
$A_{\rm Fe}$ & $6.6^{+0.7}_{-0.7}$ & $7.7^{+0.5}_{-0.6}$ & $6.0^{+0.5}_{-0.5}$ & $6.4^{+0.5}_{-0.5}$ & $5.9^{+0.4}_{-0.4}$ & $5.8^{+0.4}_{-0.4}$ & $6.5^{+0.3}_{-0.2}$ \\
$\log n_{\rm e}$ [cm$^{-3}$] & $15^\star$ & $15^\star$ & $15^\star$ & $15^\star$ & $15^\star$ & $19^\star$ & $15^\star$ \\
$\alpha_{13}$ & $-0.012^{+0.009}_{-0.018}$ &$-0.010^{+0.008}_{-0.018}$ & $-0.054^{+0.023}_{-0.028}$ & $-0.026^{+0.025}_{-0.050}$ & $-0.051^{+0.023}_{-0.024}$ & $-0.098^{+0.024}_{-0.025}$ & $-0.140^{+0.024}_{-0.051}$ \\
\hline
{\tt xillver} &&&&&&& \\
norm & -- & -- & -- & -- & -- & -- & $< 10^{-4}$ \\
\hline
$C_{\rm FPMA}$ & $1^\star$ & $1^\star$ & $1^\star$ & $1^\star$ & $1^\star$ & $1^\star$ & $1^\star$ \\ 
$C_{\rm FPMB}$ & $1.001_{-0.001}^{+0.001}$ & $1.001_{-0.001}^{+0.001}$ & $1.001_{-0.001}^{+0.001}$ & $1.001_{-0.001}^{+0.001}$ & $1.001_{-0.001}^{+0.001}$ & $1.001_{-0.001}^{+0.001}$ & $1.001_{-0.001}^{+0.001}$ \\
$C_{\rm XRT}$ & $0.955_{-0.008}^{+0.008}$ & $0.956_{-0.008}^{+0.009}$ & $0.953_{-0.023}^{+0.014}$ & $0.958_{-0.012}^{+0.014}$& $0.958_{-0.009}^{+0.014}$& $0.960_{-0.009}^{+0.013}$& $0.956_{-0.008}^{+0.008}$ \\
\hline\hline
$\chi^2/\nu$ & 869.29/647 & 874.25/645 & 869.32/647 & 872.36/647 & 868.98/646 & 874.75/647 & 870.56/646 \\
& = 1.34357 & = 1.35543 & = 1.34362 & = 1.34832 & = 1.34517 & = 1.35201 & = 1.34762 \\
\hline\hline
\end{tabular}
}
\vspace{0.3cm}
\caption{\rm Summary of the best-fit values for Models~B1-B7 after the MCMC runs. Note that the reported uncertainties correspond to 1-$\sigma$ limit for one relevant parameter ($\Delta\chi^2 = 1$). $R_{\rm in} = -1$ means that $R_{\rm in}$ is set at the ISCO radius. $^\star$ indicates that the parameter is frozen in the fit. \label{t-fit}}
\end{table*}

%%%%%%%%%%%%%%%%%%%%%%%%%%%%%%%

\section{Discussion}\label{s-dis}

Our results are in general consistent with the analysis in \citet{2016ApJ...821L...6P} and depend only weakly on the choice of the emissivity profile. We only note some minor discrepancy in the spin parameter estimate, $a_* = 0.95_{-0.08}^{+0.02}$ (1-$\sigma$) in \citet{2016ApJ...821L...6P} vs $a_* = 0.994^{+0.001}_{-0.002}$ for our Model~B1, which may be attributed to a different data reduction (with different calibration files), different versions of {\tt relxill}, and different versions of the three FITS files used by {\tt relxill} for, respectively, the reflection spectrum in the rest-frame of the gas, the transfer function, and the lamppost emissivity profile (we use here the latest versions and they should be all more accurate). Model~B1 for the lamppost geometry provides a slightly better fit than Model~B2 with the broken power-law emissivity profile, but the difference is small and the estimates of the model parameters are consistent.

From Eq.~(\ref{eq-a13}), we see that we do not recover the Kerr metric $\alpha_{13} = 0$ at the 90\% confidence level for Model~B1. However, our uncertainties only include the statistical uncertainties and, thanks to the high-quality of these data, the latter is very small. We are completely ignoring modeling uncertainties related to the theoretical model and instrumental effects~\citep[see, e.g.,][]{2020arXiv201104792B}. In the rest of this section, we will investigate some modeling uncertainties.

The choice of the emissivity profile has definitively some impact on the measurement of $\alpha_{13}$, as we can see from the (small) difference between Model~B1 and Model~B2. Even the lamppost geometry of Model~B1 is presumably an approximation of the actual coronal geometry of the black hole in GX~339--4, but the fact that Model~B1 and Model~B2 provide quite similar and consistent estimates of $\alpha_{13}$ may suggest that the choice of the emissivity profile should not induce too large systematic uncertainties on our measurement of $\alpha_{13}$.

In the fits, we set the color factor in {\tt nkbb} to $f_{\rm col} = 1.7$. The color factor is a phenomenological parameter to include the impact of non-thermal effects (mainly photon-electron scattering in the disk's atmosphere) in the thermal spectrum of the accretion disk. 1.7 is the most widely used value for a 10~$M_\odot$ black hole accreting at 10\% of its Eddington limit~\citep{1995ApJ...445..780S} and is also the value employed in \citet{2016ApJ...821L...6P}. However, in general one may expect that for a stellar-mass black hole with a thin accretion disk the color factor is in the range 1.5-1.9~\citep{1995ApJ...445..780S,2006ApJS..164..530D}. We repeat the MCMC analysis for the lamppost model for $f_{\rm col} = 1.5$ (Model~B3) and $f_{\rm col} = 1.9$ (Model~B4). The best-fit values are shown in Tab.~\ref{t-fit} (fourth and fifth column, respectively). As can be seen, even the exact choice of the color factor has a moderate impact on the final measurement of $\alpha_{13}$ and of the other parameters.

The Eddington mass accretion rate can be written as
\be
\dot{M}_{\rm Edd} = 4.7 \cdot 10^{18} \left(\frac{0.3}{\eta}\right) 
\left(\frac{M}{10 \, M_\odot}\right) \,\, \text{g/s} \, ,
\ee
where $\eta$ is the radiative efficiency and $\eta = 0.3$ for a Novikov-Thorne accretion disk of a Kerr black hole with spin parameter $a_* = 0.996$~\cite[see, e.g.,][]{2012PhRvD..85d3001B}. From Model~B1, we infer that the black hole in GX~339--4 was accreting at about 10\% of its Eddington limit, which is nicely in the 5\% to 30\% range required by a Novikov-Thorne disk with inner edge at the ISCO radius~\citep{2010ApJ...718L.117S,2011MNRAS.414.1183K,2014SSRv..183..295M}. Despite that, we can relax the assumption $R_{\rm in}=R_{\rm ISCO}$ and run another MCMC analysis. This is Model~B5 in Tab.~\ref{t-fit}, where $R_{\rm in}$ free in the fit. Model~B5 has $\chi^2$ slightly lower than Model~B1, but with one more free parameter in the fit, and the reduced $\chi^2$ is indeed a bit higher than Model~B1.

Our reflection model assumes that the disk's electron density is $n_{\rm e} = 10^{15}$~cm$^{-3}$. However, such a value is probably too low for an accreting stellar-mass black hole in an X-ray binary system and recently there have been some studies (within the Kerr framework) on the impact of the disk's electron density on the estimate of the properties of a black hole from the analysis of its reflection features~\citep{2019MNRAS.484.1972J,2019MNRAS.489.3436J}. To estimate the impact of the disk's electron density, we set $n_{\rm e} = 10^{19}$~cm$^{-3}$, which is the maximum value allowed by our {\tt xillver}-based model~\citep{2010ApJ...718..695G} but may actually be still a few orders of magnitude too low for GX~339--4~\citep{2019MNRAS.484.1972J,2019MNRAS.489.3436J}. This is our Model~B6 and its best-fit values can be seen in Tab.~\ref{t-fit} (seventh column). The disk's electron density has some impact on the estimate of $\alpha_{13}$, and in this case we move away from the Kerr limit.

\begin{figure*}[t]
\begin{center}
\includegraphics[width=5.9cm,trim={0cm 0cm 0cm 0cm},clip]{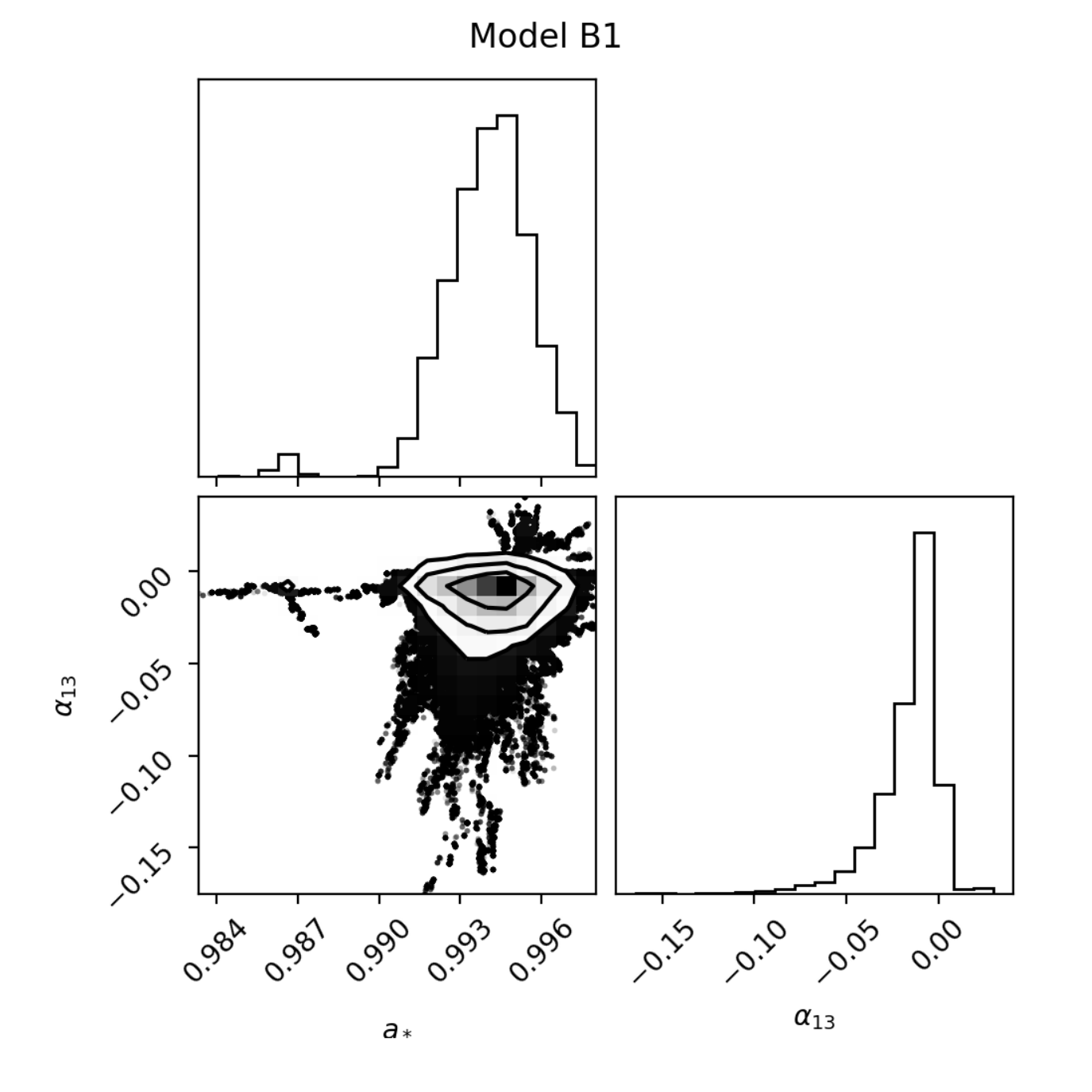}
\includegraphics[width=5.9cm,trim={0cm 0cm 0cm 0cm},clip]{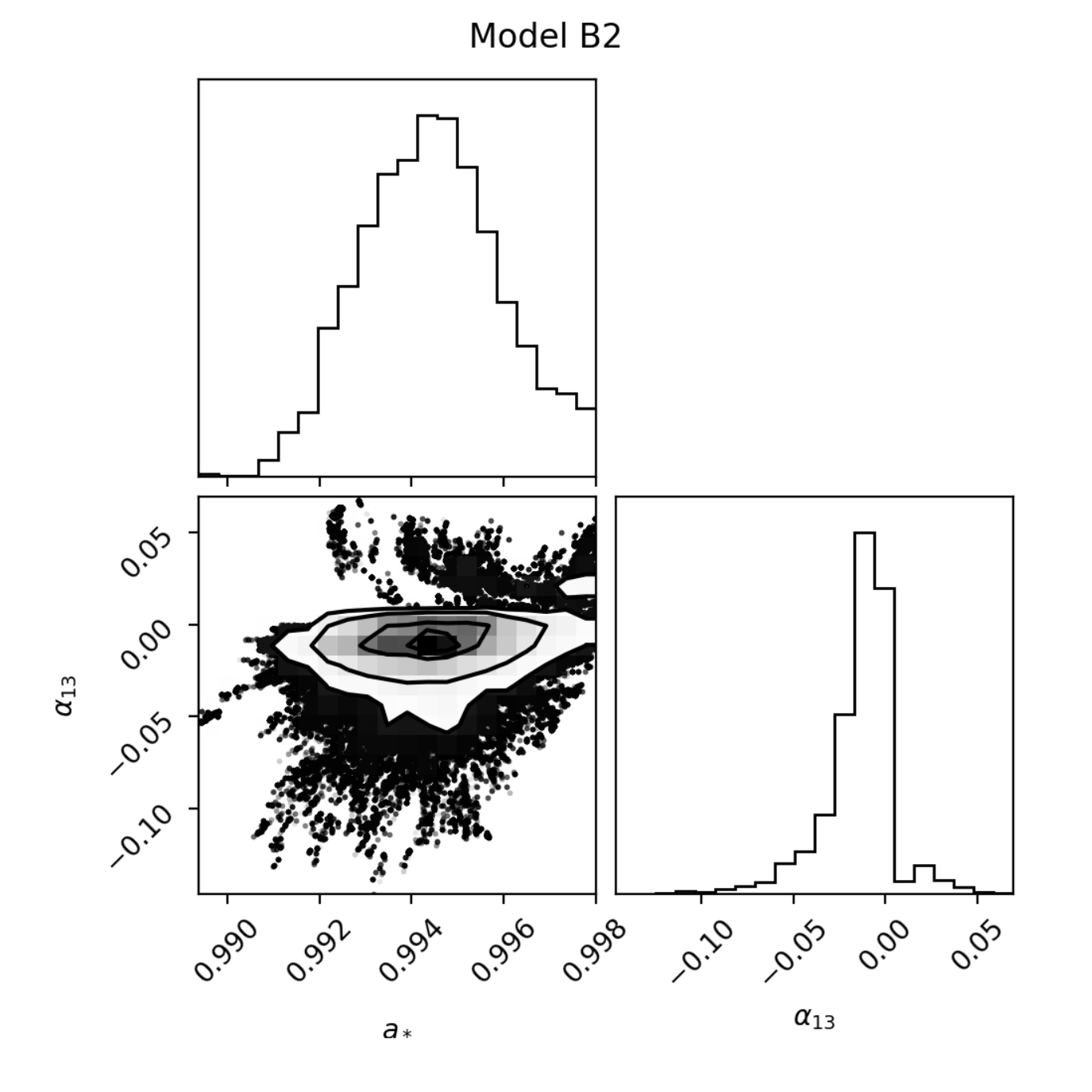} \\
\includegraphics[width=5.9cm,trim={0cm 0cm 0cm 0cm},clip]{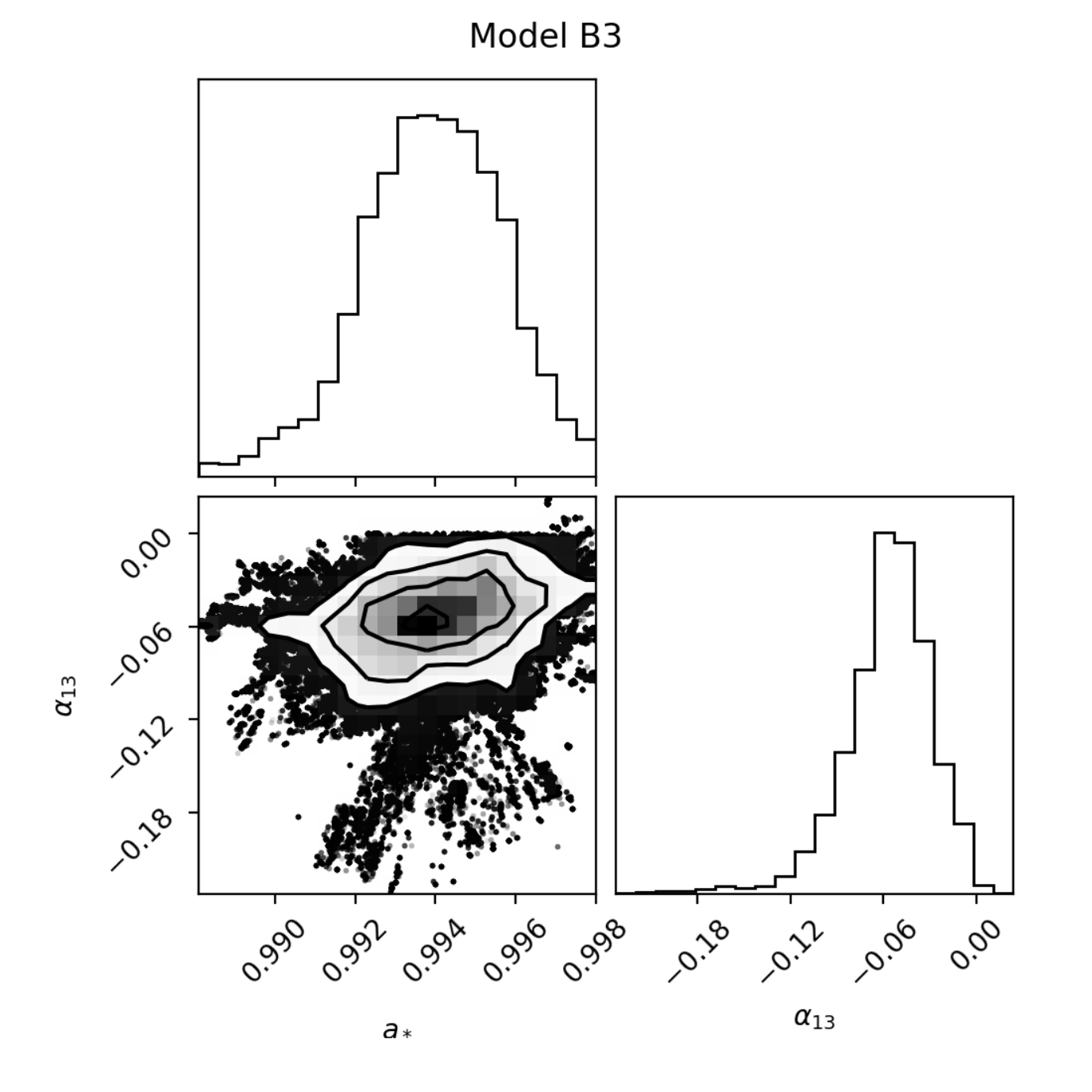}
\includegraphics[width=5.9cm,trim={0cm 0cm 0cm 0cm},clip]{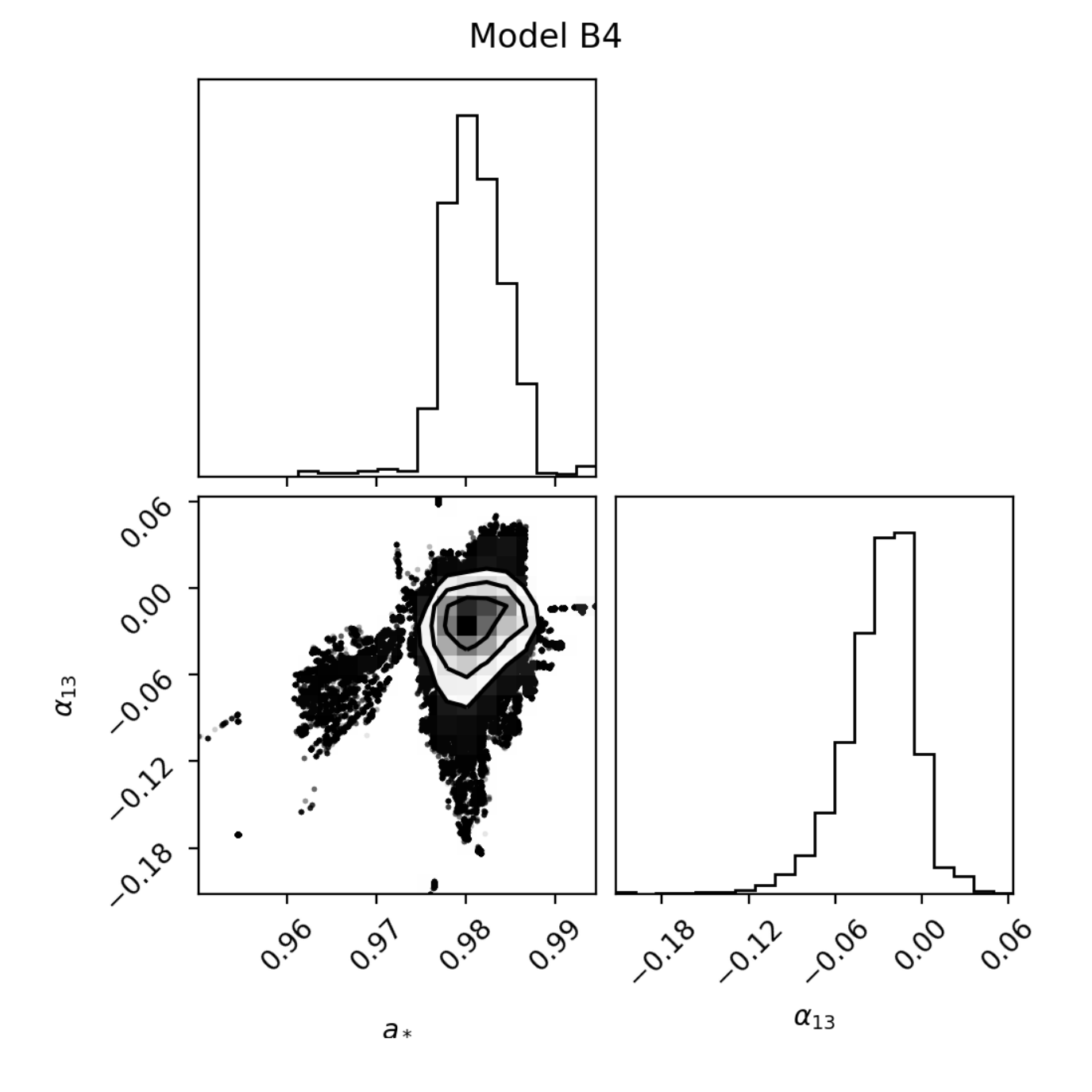} \\
\includegraphics[width=5.9cm,trim={0cm 0cm 0cm 0cm},clip]{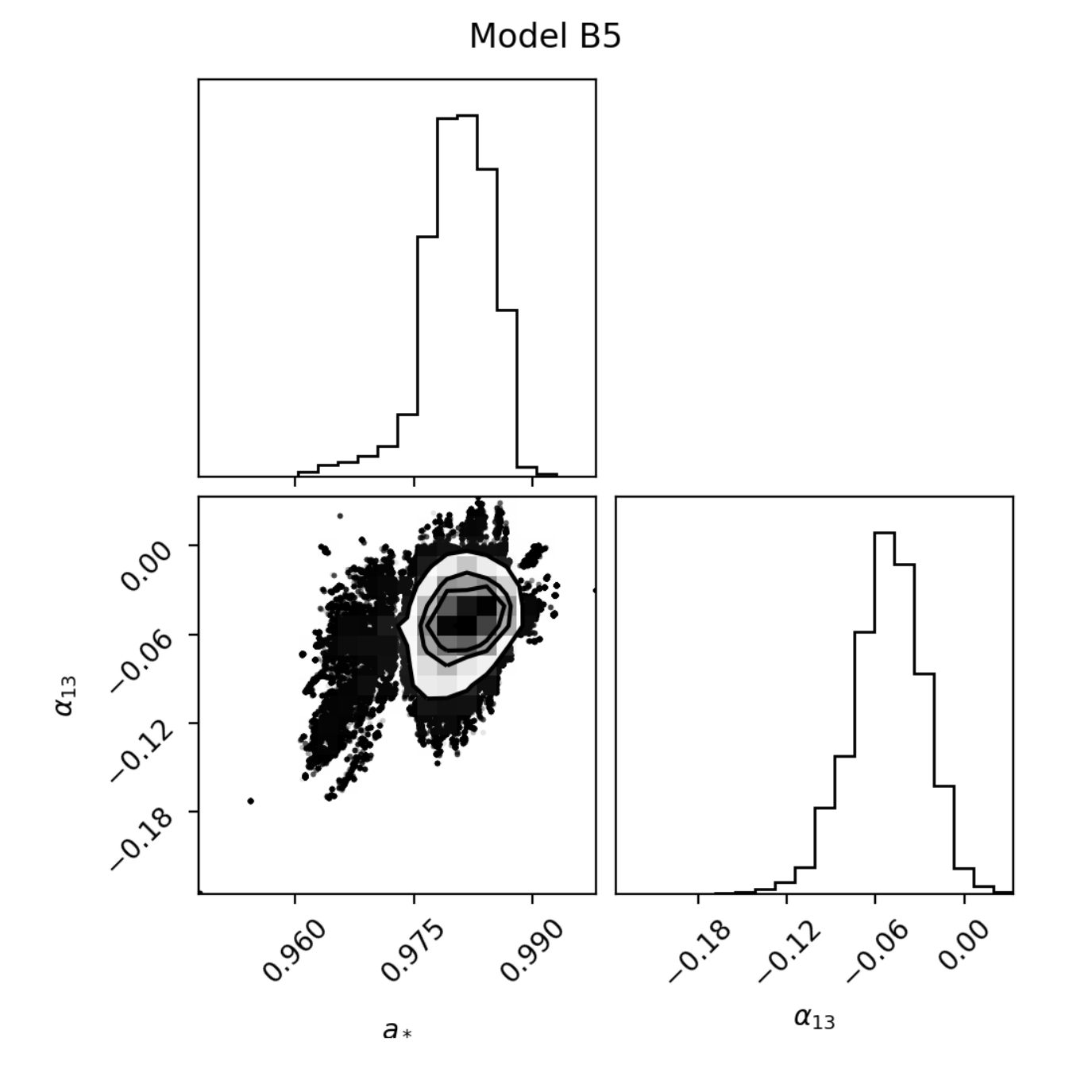}
\includegraphics[width=5.9cm,trim={0cm 0cm 0cm 0cm},clip]{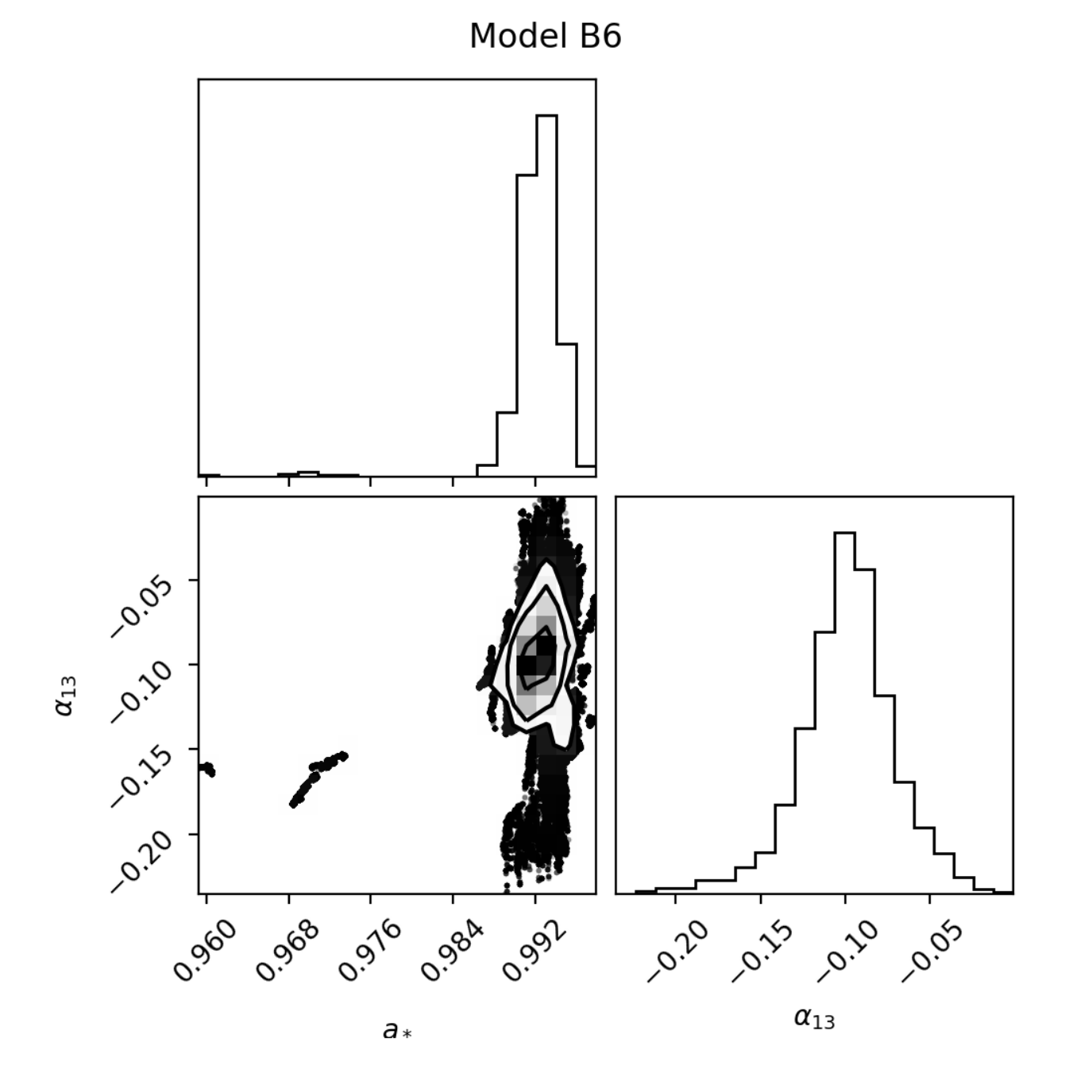}
\includegraphics[width=5.9cm,trim={0cm 0cm 0cm 0cm},clip]{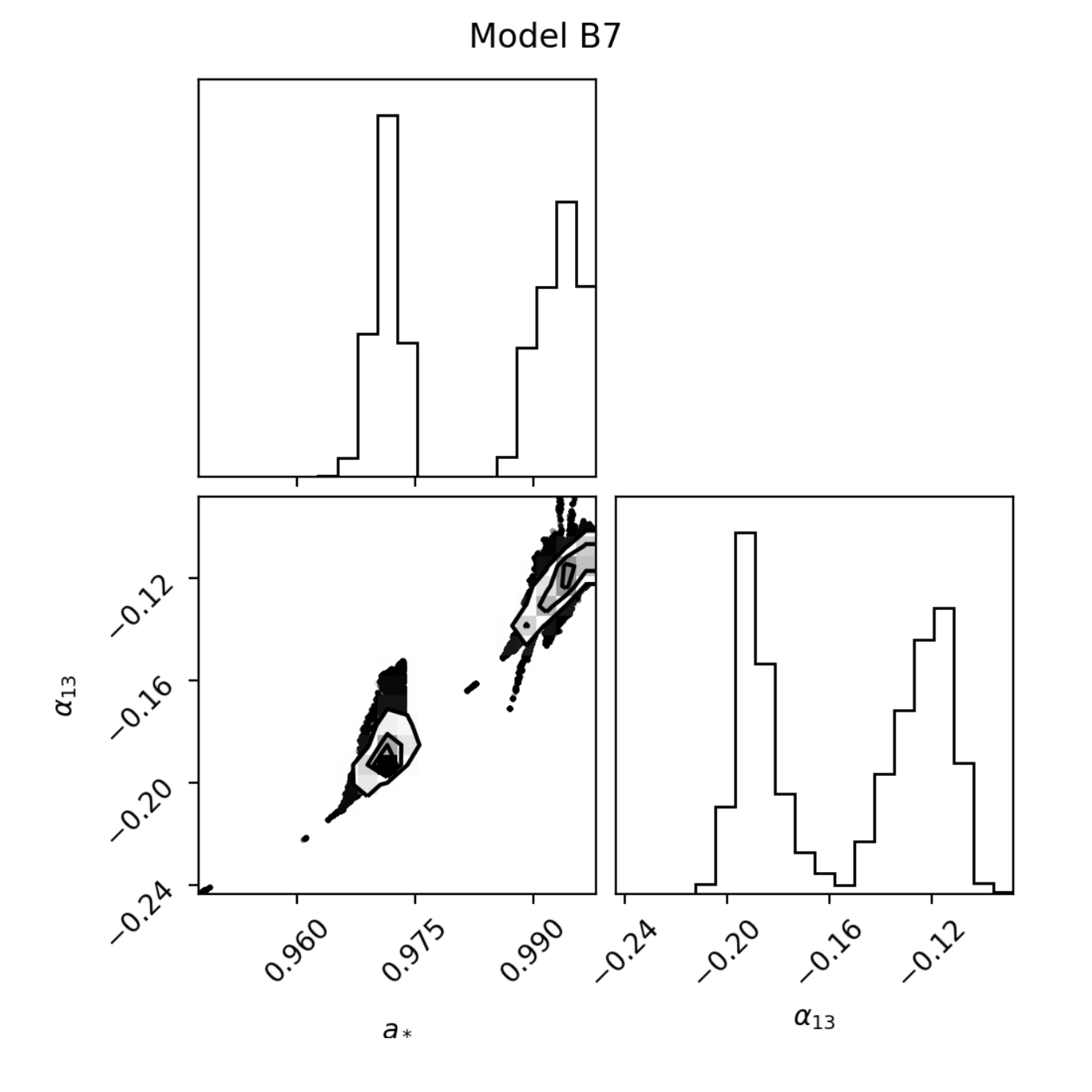} \\
\end{center}
\vspace{-0.3cm}
\caption{Constraints on the black hole spin parameter $a_*$ and the Johannsen deformation parameter $\alpha_{13}$ for Models~B1-B7 after the MCMC runs. The 2D plots report the 1-, 2-, and 3-$\sigma$ confidence contours (corresponding, respectively, to $\Delta\chi^2 = 2.3$, 6.2, and 11.8). 
\label{f-aa}}
\end{figure*}

Last, we can consider the possibility that there is some cold material nearby the source and this generates a non-relativistic reflection spectrum, which we model here with {\tt xillver} assuming that the ionization parameter is $\log\xi = 0$ (cold material). This is quite a common case among AGNs and, at some level, even for black hole binaries. The results of this fit are reported in the last column of Tab.~\ref{t-fit} under the name Model~B7. Apart from the residuals we do not see any need for an additional non-relativistic reflection spectrum, as such a component, if present, is very weak (we get only an upper bound on the value of its normalization). We note that with Model~B7 we get multiple $\chi^2$ minima and thus multiple measurements of $\alpha_{13}$ (see the islands in the Model~B7 panel in Fig.~\ref{f-aa}).

Fig.~\ref{f-aa} shows the constraints on the black hole spin parameter $a_*$ and the Johannsen deformation parameter $\alpha_{13}$ for Models~B1-B7 after the MCMC runs. If we ignore Model~B7 with multiple measurements, we 90\% confidence level window of the value of $\alpha_{13}$ provide by all other fits is
\be\label{eq-a13-tot}
-0.12 < \alpha_{13} < 0.03 \, .
\ee 
This can be considered our final measurement that includes the main modeling uncertainties.

We note that we can expect a few more modeling uncertainties that we are not able to estimate at the moment. The reflection spectrum at the emission point is calculated with {\tt xillver}, and we know that {\tt xillver} is recommended for cold disks, so either AGNs or black hole binaries in the hard state with a low mass accretion rate, while in our data we see a prominent thermal component. Previous analysis of a source in the thermal state with {\tt relxill\_nk} had already provided unclear results and we were not able to confirm the Kerr nature of the source~\citep{2019PhRvD..99l3007L}. Another simplification of the model, which may be responsible for some bias in the measurement of $\alpha_{13}$, is related to the returning radiation, which is completely ignored in our model but it can be important when the inner edge of the accretion disk is very close to the black hole (which is the case here). At the moment, there are only some partial studies on the role of the returning radiation on the estimation of the model parameters in X-ray reflection spectroscopy~\cite[see, for instance,][]{2002MNRAS.336..315R,2006A&A...453..773S,2016ApJ...821L...1N,2020MNRAS.498.3302W,2020arXiv200615838R}, but it is surely something worth exploring in the future. Considering that all fits suggest quite a small plunging region, a large radiative efficiency in the disk, and a small viewing angle, the effects of the radiation emitted from the plunging region~\citep{2020PhRvD.101l3014C} and the finite thickness of the disk~\citep{2020ApJ...899...80A} should be completely negligible.

%%%%%%%%%%%%%%%%%%%%%%%%%%%%%%%

\section{Concluding remarks}\label{s-con}

\citet{2020JCAP...05..026W} analyzed with an older version of {\tt relxill\_nk} a composite \textsl{RXTE} spectrum of GX~339--4 resulting from 23~individual exposures of the 2002 outburst. With a total exposure time of about 46~ks, the total number of photon count was around 40~million. The final estimate of the Johannsen deformation parameter was $\alpha_{13} = -0.8_{-0.6}^{+0.8}$ (90\% confidence level), but the exact choice of the model turned out to be very important on the final estimate of $\alpha_{13}$ and the selection of the model was challenging, in part because of the low energy resolution near the iron line.

In the present paper, we analyzed with {\tt nkbb} and {\tt relxill\_nk} a 30~ks \textsl{NuSTAR} observation with a 2~ks snapshot of \textsl{Swift}. The measurement of the Johannsen deformation parameter $\alpha_{13}$ is definitively more precise. From Model~B1, we get $\alpha_{13} = -0.012_{-0.039}^{+0.011}$ (90\% confidence level). The reported uncertainty corresponds to the statistical error only. Such a precision is better than the currently most stringent and reliable constraints on the Johannsen deformation parameter $\alpha_{13}$: $\alpha_{13} = 0.00_{-0.15}^{+0.05}$ obtained from the analysis of a \textsl{Suzaku} observation of the black hole binary GRS~1915+105~\citep{2020ApJ...899...80A} and $\alpha_{13} = 0.00_{-0.20}^{+0.07}$ obtained from six simultaneous observations of \textsl{NuSTAR} and \textsl{XMM-Newton} of the Seyfert~1 galaxy MCG--6--30--15~\citep{2019ApJ...875...56T}. Here all uncertainties are at 90\% confidence level for one relevant parameter ($\Delta\chi^2 = 2.71$).

We have explored several models to estimate systematic uncertainties related to the model. If we combine all the 90\% confidence level windows to have the largest possible range of $\alpha_{13}$ as a simple estimate of the modeling uncertainties, we get $-0.12 < \alpha_{13} < 0.03$, which is still a bit tighter of a constraint than previous measurements that only included statistical uncertainties.

We note that, as of now, {\tt nkbb} was only used to analyze 17~\textsl{RXTE} observations of the black hole binary LMC~X-1, finding  $\alpha_{13} = 0.32_{-3.1}^{+0.04}$ (90\% confidence level)~\citep{2020ApJ...897...84T}. Such a constraint is definitively worse than the best constraints inferred with {\tt relxill\_nk}. The main reason is that the continuum-fitting method is, in general, less powerful than X-ray reflection spectroscopy to test the Kerr nature of a source, mainly because the thermal spectrum has quite a simple shape and there is a strong degeneracy between the black hole spin parameter and the Johannsen deformation parameter $\alpha_{13}$, so it would be possible to get a very precise measurement of $\alpha_{13}$ in the presence of an independent and precise estimate of the spin parameter $a_*$, but otherwise the constraints on $a_*$ and $\alpha_{13}$ are quite weak; see Figs.~1 and 2 in \citet{2020ApJ...897...84T}.

The simultaneous analysis with {\tt nkbb} and {\tt relxill\_nk} of the present work, together with the high quality of the \textsl{NuSTAR} and \textsl{Swift} data, allows for a very precise measurement of $\alpha_{13}$, and presumably, this time, the systematic uncertainties are comparable with the statistical uncertainties. We note that we fit many parameters, including the black hole mass $M$ and distance $D$, which, in the traditional continuum-fitting method, are inferred from other observations. The accuracy of our current estimate of $\alpha_{13}$ would surely benefit from independent estimates of $M$ and $D$, which are not impossible in the future. A higher energy resolution near the iron line would also be helpful for our measurement, as that of \textsl{NuSTAR} is around 400~eV at 6~keV and we know that the iron line is the most informative feature about the spacetime metric in the strong gravity region.

Last, the constraints of this paper and of previous studies using {\tt relxill\_nk} can also be compared with the constraints that can be currently obtained from other electromagnetic tests. The 2017 Event Horizon Telescope observation of M87* provides the constraint $-3.6 < \alpha_{13} < 5.9$~\citep{2020arXiv201001055P}, which is more than an order of magnitude weaker than the constraints possible with X-ray reflection spectroscopy.

%%%%%%%%%%%%%%%%%%%%%%%%%%%%%%%

\vspace{0.5cm}

{\bf Acknowledgments --}
This work was supported by the Innovation Program of the Shanghai Municipal Education Commission, Grant No.~2019-01-07-00-07-E00035, the National Natural Science Foundation of China (NSFC), Grant No.~11973019, and Fudan University, Grant No.~JIH1512604. V.G. is supported through the Margarete von Wrangell fellowship by the ESF and the Ministry of Science, Research and the Arts Baden-W\"urttemberg. D.A. is supported through the Teach@T\"ubingen Fellowship.
A.T., C.B., and V.G. are members of the International Team~458 at the International Space Science Institute (ISSI), Bern, Switzerland, and acknowledge support from ISSI during the meetings in Bern.

%%%%%%%%%%%%%%%%%%%%%%%%%%%%%%%

\appendix

\section{Johannsen spacetime}\label{appendix}

The Johannsen metric is one of the so-called parametric black hole spacetimes proposed in literature and specifically designed to test the Kerr black hole hypothesis with electromagnetic observations of  black holes~\citep{2013PhRvD..88d4002J}. In Boyer-Lindquist-like coordinates, its line element reads~\citep{2013PhRvD..88d4002J}
\be\label{eq-jm}
ds^2 &=&-\frac{\tilde{\Sigma}\left(\Delta-a^2A_2^2\sin^2\theta\right)}{B^2}dt^2
+\frac{\tilde{\Sigma}}{\Delta A_5}dr^2+\tilde{\Sigma} d\theta^2 
-\frac{2a\left[\left(r^2+a^2\right)A_1A_2-\Delta\right]\tilde{\Sigma}\sin^2\theta}{B^2}dtd\phi \nonumber\\
&&+\frac{\left[\left(r^2+a^2\right)^2A_1^2-a^2\Delta\sin^2\theta\right]\tilde{\Sigma}\sin^2\theta}{B^2}d\phi^2
\ee
where $M$ is the black hole mass, $a = J/M$, $J$ is the black hole spin angular momentum, $\tilde{\Sigma} = \Sigma + f$, and
\be
\Sigma = r^2 + a^2 \cos^2\theta \, , \qquad
\Delta = r^2 - 2 M r + a^2 \, , \qquad
B = \left(r^2+a^2\right)A_1-a^2A_2\sin^2\theta \, .
\ee
The functions $f$, $A_1$, $A_2$, and $A_5$ are defined as
\be
f = \sum^\infty_{n=3} \epsilon_n \frac{M^n}{r^{n-2}} \, , \quad
A_1 = 1 + \sum^\infty_{n=3} \alpha_{1n} \left(\frac{M}{r}\right)^n \, , \quad
A_2 = 1 + \sum^\infty_{n=2} \alpha_{2n}\left(\frac{M}{r}\right)^n \, , \quad
A_5 = 1 + \sum^\infty_{n=2} \alpha_{5n}\left(\frac{M}{r}\right)^n \, ,
\ee
where $\{ \epsilon_n \}$, $\{ \alpha_{1n} \}$, $\{ \alpha_{2n} \}$, and $\{ \alpha_{5n} \}$ are four infinite sets of deformation parameters without constraints from the Newtonian limit and Solar System experiments. $\epsilon_3$, $\alpha_{13}$, $\alpha_{22}$, and $\alpha_{52}$ are the leading order deformation parameters. In this work, for the sake of simplicity, we have restricted our attention only to deformation parameter $\alpha_{13}$, which has the strongest impact on the reflection spectrum. However, our analysis can be easily extended to metrics with other non-vanishing deformation parameters as well as, more in general, to any stationary, axisymmetric, and asymptotically flat black hole spacetime.

In order to avoid pathological properties in the spacetime, we need to impose some constraints on the values of $a_*$ and $\alpha_{13}$. For the spin parameter, we require $| a_* | \le 1$. As in the Kerr spacetime, for $| a_* | > 1$ there is no event horizon and the metric describes the spacetime of a naked singularity. For the deformation parameter $\alpha_{13}$, we require that~\citep[for the details, see][]{2018PhRvD..98b3018T}
\be
\label{eq-constraints}
\alpha_{13} > - \frac{1}{2} \left( 1 + \sqrt{1 - a^2_*} \right)^4 \, .
\ee

%%%%%%%%%%%%%%%%%%%%%%%%%%%%%%%

\end{document}